\documentstyle[12pt]{article}
\newcommand{\Cslash}{\not \!\! C}
\newcommand{\Dslash}{\not \!\! D}
\newcommand{\Aslash}{\not \!\! A}
\newcommand{\kslash}{\not \!\! k}
\newcommand{\delslash}{\not \! \partial}
\begin{document}

\begin{flushright}{UT-884\\
April 2000
}
\end{flushright}
\vskip 0.5 truecm

\begin{center}
{\Large{\bf Algebraic Generalization of the Ginsparg-Wilson
Relation }}
\end{center}
\vskip .5 truecm
\centerline{\bf Kazuo Fujikawa}
\vskip .4 truecm
\centerline {\it Department of Physics,University of Tokyo}
\centerline {\it Bunkyo-ku,Tokyo 113,Japan}
\vskip 0.5 truecm

\makeatletter
\@addtoreset{equation}{section}
\def\theequation{\thesection.\arabic{equation}}
\makeatother

\begin{abstract}
A specific algebraic realization of the Ginsparg-Wilson relation 
in the form $\gamma_{5}(\gamma_{5}D)+(\gamma_{5}D)\gamma_{5} =
 2a^{2k+1}(\gamma_{5}D)^{2k+2}$ 
is discussed, where $k$ stands for a non-negative integer and 
$k=0$ corresponds to the commonly discussed Ginsparg-Wilson 
relation. From a view point of algebra, a characteristic 
property of our proposal is that we have a closed algebraic 
relation for one unknown operator $D$, although this relation 
itself is obtained from the original proposal of 
Ginsparg and Wilson, $\gamma_{5}D+D\gamma_{5}=2aD\gamma_{5}
\alpha D$, by choosing $\alpha$ as an operator containing
$D$ ( and thus Dirac matrices).
In this paper, it is shown that we can construct the operator 
$D$ explicitly for any value of $k$. We first show that
the instanton-related index of all these operators is identical. 
We then illustrate in detail a generalization of Neuberger's 
overlap Dirac operator to the case  $k=1$. On the basis of 
explicit construction, it is shown that the chiral 
symmetry breaking term becomes more irrelevent for larger 
$k$ in the sense of Wilsonian renormalization group.
We thus have an infinite tower of new lattice Dirac operators 
which are topologically proper, but a large enough lattice is 
required to accomodate a Dirac operator with a large 
value of $k$.
\end{abstract}

\section{Introduction}
Recent developments in the treatment of  fermions in lattice 
gauge theory are based on a  hermitian lattice Dirac  operator 
$\gamma_{5}D$ which satisfies the Ginsparg-Wilson relation[1]
\footnote{To be precise, the general relation 
$\gamma_{5}D + D\gamma_{5}=2aD\gamma_{5}\alpha D$, where 
$\alpha$ is a local operator, has been proposed in Ref.[1], 
although the authors in Ref.[1] 
analyzed ``only the simplest case where the matrix $\alpha$ is 
proportional to the unit matrix in Dirac space''. With this 
qualification in mind, we refer to (1.1) as the ``ordinary 
Ginsparg-Wilson relation'' in this paper. The original Ginsparg-
Wilson relation is more general as stated above.}
\begin{equation}
\gamma_{5}D + D\gamma_{5} = 2aD\gamma_{5}D
\end{equation}
where the lattice spacing $a$ is utilized to make a dimensional
consideration transparent, and 
$\gamma_{5}$ is a hermitian chiral Dirac matrix. 
An explicit example of the operator satisfying (1.1) and free of 
species doubling has been given by Neuberger[2].  The relation 
(1.1) led to an interesting analysis of the notion of index in 
lattice gauge theory[3]. This index theorem in turn led to a 
new form of chiral symmetry, and the chiral anomaly is obtained 
as a non-trivial Jacobian factor under this modified chiral
 transformation[4]. This chiral Jacobian is regarded as 
a lattice 
generalization of the continuum path integral[5]. The very 
detailed analyses of the lattice chiral Jacobian have been 
performed[6]-[8].
It is also possible to formulate the lattice index theorem in 
a manner[9] analogous to the continuum index theorem[10][11]. 
An interesting chirality sum rule, which relates the number of 
zero modes to that of the heaviest states, has also been 
noticed[12]. 

In this paper we discuss a generalization of the 
relation (1.1), which is characterized by a non-negative 
integer $k$. It is shown that the explicit construction of
 an infinite tower of lattice Dirac operators which satisfy the 
index theorem is possible, but a 
large enough lattice is required to accomodate a Dirac 
operator with a large value of $k$.

\section{Generalized algebra and its representation}

We discuss a generalization of the algebra (1.1) to
the form\footnote{This relation is obtained from the proposal 
in Ref. [1],$\gamma_{5}D + D\gamma_{5}
=2aD\gamma_{5}\alpha D$, by choosing $\alpha$ as an operator 
containing $D$ itself (and thus Dirac matrices).
From a view point of algebra, the original construction in [1] 
contains two unknown operators and one relation. In our 
construction, we have a closed algebraic relation for one 
unknown operator $D$, which allows a neat analyis of 
representation in this Section. This specific algebraically closed
 realization ,which is characterized by a non-negative integer, 
has not been discussed in Ref.[1].}
\begin{equation}
\gamma_{5}(\gamma_{5}D)+(\gamma_{5}D)\gamma_{5}=
2a^{2k+1}(\gamma_{5}D)^{2k+2}
\end{equation}
where $k$ stands for a non-negative integer and $k=0$ corresponds
to the ordinary Ginsparg-Wilson relation. 
 When one defines 
\begin{equation}
H\equiv \gamma_{5}aD
\end{equation}
(2.1) is rewritten as 
\begin{equation}
\gamma_{5}H+H\gamma_{5}=2H^{2k+2}
\end{equation}
or equivalently
\begin{equation}
\Gamma_{5}H+\Gamma_{5}H=0
\end{equation}
where we defined 
\begin{equation}
\Gamma_{5}\equiv \gamma_{5}-H^{2k+1}.
\end{equation}
Note that both of $H$ and $\Gamma_{5}$ are hermitian operators.

We now discuss a general representation of the algebraic relation
(2.4) following the analysis in Appendix of Ref.[13].(In Ref.[13],
the algebra was normalized as 
$\gamma_{5}(\gamma_{5}D)+(\gamma_{5}D)\gamma_{5}=
a(\gamma_{5}D)^{2}$, but here we use the normalization (2.1) to 
simplify various expressions.) 
The relation (2.4) suggests that if
\begin{equation}
H\phi_{n} = a\lambda_{n}\phi_{n}, \ \ \ (\phi_{n},\phi_{n})=1 
\end{equation}
with a real eigenvalue $a\lambda_{n}$ for the hermitian 
operator $H$, then
\begin{equation}
H(\Gamma_{5}\phi_{n}) = -a\lambda_{n}(\Gamma_{5}\phi_{n}).
\end{equation}
Namely, the eigenvalues $\lambda_{n}$ and $-\lambda_{n}$ are 
always paired if $\lambda_{n}\neq 0$ and 
$(\Gamma_{5}\phi_{n},\Gamma_{5}\phi_{n})\neq 0$.
We also note the relation, which is derived by sandwiching the
relation (2.3) by $\phi_{n}$,
\begin{equation}
(\phi_{n},\gamma_{5}\phi_{n})=(a\lambda_{n})^{2k+1}\ \ \ \ for\ \
 \ \lambda_{n}\neq 0.
\end{equation}
Consequently
\begin{equation}
|(a\lambda_{n})^{2k+1}|= |(\phi_{n},\gamma_{5}\phi_{n})|\leq
||\phi_{n}||||\gamma_{5}\phi_{n}||=1.
\end{equation}
Namely, all the possible eigenvalues are bounded by
\begin{equation}
|\lambda_{n}|\leq \frac{1}{a}.
\end{equation}

We thus  evaluate the norm of $\Gamma_{5}\phi_{n}$
\begin{eqnarray}
(\Gamma_{5}\phi_{n},\Gamma_{5}\phi_{n})
&=& (\phi_{n},(\gamma_{5}-H^{2k+1})(\gamma_{5}-H^{2k+1})\phi_{n})
\nonumber\\ 
&=&(\phi_{n},(1-H^{2k+1}\gamma_{5}-\gamma_{5}H^{2k+1}+H^{2(2k+1)})
\phi_{n})\nonumber\\
&=&[1-(a\lambda_{n})^{2(2k+1)}]\nonumber\\
&=&[1-(a\lambda_{n})^{2}][1+(a\lambda_{n})^{2}+...+
(a\lambda_{n})^{4k}]
\end{eqnarray}
where we used (2.8).
By remembering that all the eigenvalues are real, we find that 
$\phi_{n}$ is a ``highest'' state 
\begin{equation}
\Gamma_{5}\phi_{n}=0
\end{equation}
only if 
\begin{equation}
[1-(a\lambda_{n})^{2}]=(1-a\lambda_{n})(1+a\lambda_{n})=0 
\end{equation}
for the Euclidean positive definite inner 
product $(\phi_{n}, \phi_{n})\equiv\sum_{x}\phi_{n}^{\dagger}(x)
\phi_{n}(x)$.\\
We thus conclude that the states $\phi_{n}$ with 
$\lambda_{n}= \pm \frac{1}{a}$
 are {\em not} paired by the operation $\Gamma_{5}\phi_{n}$ and 
\begin{equation}
\gamma_{5}D\phi_{n}= \pm \frac{1}{a}\phi_{n}, \ \ \ \gamma_{5}
\phi_{n}= \pm \phi_{n}
\end{equation}
respectively. These eigenvalues are in fact the maximum or minimum
of the possible eigenvalues of $H/a$ due to (2.10).

As for the vanishing eigenvalues $H\phi_{n}=0$, we find from
(2.4) that $H\gamma_{5}\phi_{n}=0$, namely, 
$H[(1\pm\gamma_{5})/2]\phi_{n}=0$. We thus have 
\begin{equation}
\gamma_{5}D\phi_{n}=0,\ \ \ \gamma_{5}\phi_{n}=\phi_{n} \ \ \ or \ \ \ 
\gamma_{5}\phi_{n}=-\phi_{n}.
\end{equation}
\\
To summarize the analyses so far, all the normalizable 
eigenstates $\phi_{n}$ of $\gamma_{5}D=H/a$ are categorized into 
the following 3 classes:\\
(i)\ $n_{\pm}$ (``zero modes''),\\
\begin{equation}
\gamma_{5}D\phi_{n}=0, \ \ \gamma_{5}\phi_{n} = \pm \phi_{n},
\end{equation}
(ii)\ $N_{\pm}$ (``highest states''), \\
\begin{equation}
\gamma_{5}D\phi_{n}= \pm \frac{1}{a}\phi_{n}, \ \ \
\gamma_{5}\phi_{n} = \pm \phi_{n},\ \ \ respectively,
\end{equation}
(iii)``paired states'' with $0 < |\lambda_{n}| < 1/a$,
\begin{equation}
\gamma_{5}D\phi_{n}= \lambda_{n}\phi_{n}, \ \ \ 
\gamma_{5}D(\Gamma_{5}\phi_{n})
= - \lambda_{n}(\Gamma_{5}\phi_{n}).
\end{equation}
Note that $\Gamma_{5}(\Gamma_{5}\phi_{n})\propto \phi_{n}$ for 
$0<|\lambda_{n}|<1/a$.\\

We thus obtain the index relation[3][4]
\begin{eqnarray}
Tr\Gamma_{5}&\equiv& \sum_{n}(\phi_{n},\Gamma_{5}\phi_{n})
\nonumber\\
&=&\sum_{ \lambda_{n}=0}(\phi_{n},\Gamma_{5}\phi_{n})
+\sum_{0<|\lambda_{n}|<1/a}(\phi_{n},\Gamma_{5}\phi_{n})
+\sum_{|\lambda_{n}|=1/a}(\phi_{n},\Gamma_{5}\phi_{n})
\nonumber\\
&=&\sum_{\lambda_{n}=0}(\phi_{n},\Gamma_{5}\phi_{n})\nonumber\\
&=&\sum_{\lambda_{n}=0}(\phi_{n},(\gamma_{5}-H^{2k+1})\phi_{n})
\nonumber\\
&=&\sum_{\lambda_{n}=0}(\phi_{n},\gamma_{5}\phi_{n})
\nonumber\\
&=& n_{+} - n_{-} =  index
\end{eqnarray}
where $n_{\pm}$ stand for the number of  normalizable zero modes
with $\gamma_{5}\phi_{n}=\pm\phi_{n}$ in the classification (i) 
above. We here used the fact that 
$\Gamma_{5}\phi_{n}=0$ for the ``highest states'' and that 
$\phi_{n}$ and $\Gamma_{5}\phi_{n}$ are orthogonal to each other
for $0<|\lambda_{n}|<1/a$ since they have eigenvalues 
with opposite signatures.

On the other hand, the relation $Tr \gamma_{5}=0$, which is
expected to be valid in (finite) lattice theory,  leads to ( by 
using (2.8))
\begin{eqnarray}
Tr\gamma_{5} &=& \sum_{n}(\phi_{n},\gamma_{5}\phi_{n})\nonumber\\
&=& \sum_{ \lambda_{n}=0}(\phi_{n},\gamma_{5}\phi_{n}) +
\sum_{\lambda_{n}\neq 0}(\phi_{n},\gamma_{5}\phi_{n})\nonumber\\
&=& n_{+} - n_{-}+\sum_{ \lambda_{n}\neq 0}(a\lambda_{n})^{2k+1}
=0.
\end{eqnarray}
In the last line  of this relation, all the states except for the 
``highest states'' with $\lambda_{n}= \pm 1/a$  cancel pairwise 
for $\lambda_{n}\neq0$. We thus obtain a chirality  sum rule[12] 
$n_{+}-n_{-}+N_{+}-N_{-}=0$ or 
\begin{equation}
n_{+}+N_{+}=n_{-}+N_{-}  
\end{equation} 
where $N_{\pm}$ stand for the number of ``highest states'' with
$\gamma_{5}\phi_{n}=\pm\phi_{n}$ in the classification (ii) 
above. These 
relations show that the chirality asymmetry at vanishing 
eigenvalues is balanced by the chirality asymmetry at the 
largest eigenvalues with $|\lambda_{n}|=1/a$. It was argued in 
Ref.[13] that $N_{\pm}$ states are the topological 
(instanton-related) excitations of the would-be species doublers.

All the $n_{\pm}$ and $N_{\pm}$ states are the eigenstates of 
$D$, $D\phi_{n}=0$ and $D\phi_{n}= (1/a)\phi_{n}$, 
respectively. If one denotes the number of states  in the 
classification (iii) above by $2N_{0}$, the total number of 
states (the dimension of the representation) $N$ is given by 
\begin{equation}
N = 2(n_{+}+N_{+}+N_{0})
\end{equation}
which is expected to be common to all the algebraic relations 
in (2.1) and to be  a constant independent of background 
gauge field configurations.

We note that all the states $\phi_{n}$ with 
$0<|\lambda_{n}|<1/a$, 
which appear pairwise with $\lambda_{n}= \pm |\lambda_{n}|$, 
can be normalized to satisfy the relations
\begin{eqnarray}
\Gamma_{5}\phi_{n}&=&
[1-(a\lambda_{n})^{2(2k+1)}]^{1/2}\phi_{-n},
\nonumber\\
\gamma_{5}\phi_{n}&=&(a\lambda_{n})^{2k+1}\phi_{n}+
[1-(a\lambda_{n})^{2(2k+1)}]^{1/2}\phi_{-n}.
\end{eqnarray}
Here $\phi_{-n}$ stands for the eigenstate with an eigenvalue
opposite to that of $\phi_{n}$.
These states $\phi_{n}$ cannot be the eigenstates of 
$\gamma_{5}$ since 
$|(\phi_{n},\gamma_{5}\phi_{n})|=|(a\lambda_{n})^{2k+1}|<1$. 

We have thus established that the representation of all the 
algebraic relations (2.1) has a similar structure. In the next
Section, we show that the index $n_{+}-n_{-}$ is identical
to all these algebraic relations if the operator $\gamma_{5}D$
satisfies suitable conditions.
 
\section{Chiral Jacobian and the index relation}

The Euclidean path integral for a fermion is defined by
\begin{equation}
\int{\cal D}\bar{\psi}{\cal D}\psi\exp[\int\bar{\psi}D\psi]
\end{equation}
where
\begin{equation}
\int\bar{\psi}D\psi\equiv \sum_{x,y}\bar{\psi}(x)D(x,y)\psi(y)
\end{equation}
and the summation runs over all the points on the lattice.
The relation (2.4) is re-written as 
\begin{equation}
\gamma_{5}\Gamma_{5}\gamma_{5}D+D\Gamma_{5}=0
\end{equation}
and thus the Euclidean action is invariant under the global
 ``chiral'' transformation[4]
\begin{eqnarray}
&&\bar{\psi}(x)\rightarrow\bar{\psi}^{\prime}(x)=
\bar{\psi}(x)+i\sum_{z}\bar{\psi}(z)\epsilon\gamma_{5}
\Gamma_{5}(z,x)\gamma_{5}
\nonumber\\
&&\psi(y)\rightarrow\psi^{\prime}(y)=
\psi(y)+i\sum_{w}\epsilon\Gamma_{5}(y,w)\psi(w)
\end{eqnarray}
with an infinitesimal constant parameter $\epsilon$.
Under this transformation, one obtains a Jacobian factor
\begin{equation}
{\cal D}\bar{\psi}^{\prime}{\cal D}\psi^{\prime}=
J{\cal D}\bar{\psi}{\cal D}\psi
\end{equation}
with
\begin{equation}
J=\exp[-2iTr\epsilon\Gamma_{5}]=\exp[-2i\epsilon(n_{+}-n_{-})]
\end{equation}
where we used the index relation (2.19).

We now relate this index appearing in the Jacobian to the 
Pontryagin index of the gauge field in a smooth continuum limit 
by following the procedure in Ref.[9].
We  start with
\begin{equation}
Tr\{\Gamma_{5}f(\frac{(\gamma_{5}D)^{2}}{M^{2}})\}
=Tr\{\Gamma_{5}f(\frac{(H/a)^{2}}{M^{2}})\}
=n_{+} - n_{-}
\end{equation}
Namely, the index is not modified by any  regulator $f(x)$ with 
$f(0)=1$ and $f(x)$ rapidly going to zero for 
$x\rightarrow\infty$, as can be confirmed by using (2.19). This
means that you can use {\em any} suitable $f(x)$ in the 
evaluation of the index by taking advantage of this property.
We then consider a local version of the index
\begin{equation}
tr\{\Gamma_{5}f(\frac{(\gamma_{5}D)^{2}}{M^{2}})\}(x,x)
=tr\{(\gamma_{5}-H^{2k+1})f(\frac{(\gamma_{5}D)^{2}}{M^{2}})\}
(x,x)
\end{equation}
where trace stands for Dirac and Yang-Mills indices; Tr in (3.7) 
includes a sum over the lattice points $x$.  
A local version of the index is not sensitive to the precise 
boundary condition , and  one may take an infinite volume 
limit of the lattice in the above expression. 

We now examine the continuum limit $a\rightarrow 0$ of the above 
local expression (3.8)\footnote{This continuum limit corresponds to the 
so-called ``naive'' continuum limit in the context of lattice gauge 
theory.}. We first observe that the term
\begin{equation}
tr\{H^{2k+1}f(\frac{(\gamma_{5}D)^{2}}{M^{2}})\}
\end{equation}
goes to zero in this limit. The large eigenvalues of 
$H=a\gamma_{5}D$ are truncated at the value $\sim aM$ by the 
regulator $f(x)$ which rapidly goes to zero for large $x$. In 
other words, the global index of the operator 
$TrH^{2k+1}f(\frac{(\gamma_{5}D)^{2}}{M^{2}})\sim O(aM)^{2k+1}$.

We thus examine the small $a$ limit of 
\begin{equation}
tr\{\gamma_{5}f(\frac{(\gamma_{5}D)^{2}}{M^{2}})\}.
\end{equation}
The operator appearing in this expression is well regularized by 
the function $f(x)$ , and we evaluate the above trace by using 
the plane wave basis to extract an explicit gauge field 
dependence.
We consider a square lattice where the momentum is defined in 
the Brillouin zone
\begin{equation}
-\frac{\pi}{2a}\leq k_{\mu} < \frac{3\pi}{2a}.
\end{equation}
We assume that the operator $D$ is free of  species doubling; in 
other words, the operator $D$ blows up rapidly 
($\sim \frac{1}{a}$) for small $a$ in the momentum region 
corresponding to species doublers. The contributions of doublers 
are eliminated by the regulator $f(x)$ in the above expression, 
since
\begin{equation}
tr\{\gamma_{5}f(\frac{(\gamma_{5}D)^{2}}{M^{2}})\}\sim
(\frac{1}{a})^{4}f(\frac{1}{(aM)^{2}})\rightarrow 0
\end{equation}
for $a\rightarrow 0$ if one chooses $f(x)=e^{-x}$, for example. 

We thus examine the above trace in the momentum range of the 
physical species
\begin{equation}
-\frac{\pi}{2a}\leq k_{\mu} < \frac{\pi}{2a}.
\end{equation}
We obtain the limiting $a\rightarrow 0$ expression
\begin{eqnarray}
&&\lim_{a\rightarrow 0}tr\{\gamma_{5}f(\frac{(\gamma_{5}D)^{2}}
{M^{2}})\}(x,x)\nonumber\\
&=& \lim_{a\rightarrow 0}tr \int_{-\frac{\pi}{2a}}^{\frac{\pi}
{2a}}\frac{d^{4}k}{(2\pi)^{4}}e^{-ikx}\gamma_{5}
f(\frac{(\gamma_{5}D)^{2}}{M^{2}})e^{ikx}\nonumber\\
&=&\lim_{L\rightarrow\infty}\lim_{a\rightarrow 0}tr 
\int_{-L}^{L}\frac{d^{4}k}{(2\pi)^{4}}e^{-ikx}\gamma_{5}
f(\frac{(\gamma_{5}D)^{2}}{M^{2}})e^{ikx}\nonumber\\
&=&\lim_{L\rightarrow\infty}tr \int_{-L}^{L}
\frac{d^{4}k}{(2\pi)^{4}}e^{-ikx}\gamma_{5}
f(\frac{(i\gamma_{5}\Dslash)^{2}}{M^{2}})e^{ikx}\nonumber\\
&\equiv&tr\{\gamma_{5}f(\frac{\Dslash^{2}}{M^{2}})\}
\end{eqnarray}
where  we first take the limit $a\rightarrow 0$ with fixed 
$k_{\mu}$ in $-L\leq k_{\mu} \leq L$, and then take the limit 
$L\rightarrow \infty$. This 
procedure is justified if the integral is well convergent
\footnote{
To be precise, we deal with an integral of the structure 
$\int^{\frac{\pi}{2a}}_{-\frac{\pi}{2a}}dx f_{a}(x)=
\int^{\frac{\pi}{2a}}_{L}dx f_{a}(x) + \int^{L}_{-L}dx f_{a}(x)
+\int^{-L}_{-\frac{\pi}{2a}}dx f_{a}(x)$
where $f_{a}(x)$ depends on the parameter $a$. (A generalization 
to a 4-dimensional integral is straightforward.) We thus have to 
prove that both of 
$\lim_{a\rightarrow 0}\int^{\frac{\pi}{2a}}_{L}dx f_{a}(x)$ and 
$\lim_{a\rightarrow 0}\int^{-L}_{-\frac{\pi}{2a}}dx f_{a}(x)$ 
can be made arbitrarily small if one lets $L$ to be large. 
A typical integral we encounter in lattice theory has a generic 
structure $\lim_{a\rightarrow 0}
\int_{-\pi/2a}^{\pi/2a} dx e^{-[\sin^{2}ax+(1-\cos 2ax)^{2}]
/(a^{2}M^{2})}=\lim_{a\rightarrow 0}
\int_{\epsilon\pi/2a}^{\pi/2a} dx 
e^{-[\sin^{2}ax+(1-\cos 2ax)^{2}]
/(a^{2}M^{2})}+\lim_{a\rightarrow 0}
\int_{-\pi/2a}^{-\epsilon\pi/2a}dx 
e^{-[\sin^{2}ax+(1-\cos 2ax)^{2}]
/(a^{2}M^{2})}+\lim_{a\rightarrow 0}
\int_{-\epsilon\pi/2a}^{\epsilon\pi/2a}dx 
e^{-[\sin^{2}ax+(1-\cos 2ax)^{2}]
/(a^{2}M^{2})}=
\lim_{a\rightarrow 0}
\int_{-\epsilon\pi/2a}^{\epsilon\pi/2a}dx 
e^{-[\sin^{2}ax+(1-\cos 2ax)^{2}]
/(a^{2}M^{2})}
=\lim_{L\rightarrow \infty}\int_{-L}^{L} dx e^{-x^{2}/M^{2}}$ 
and satisfies the above criterion, if one chooses the regulator 
$f(x)=e^{-x}$: Here $\epsilon$ is an arbitrary small fixed  
parameter, and the left-hand side of this relation stands for 
a conventional lattice calculation and the right-hand side stands
for a continuum calculation.}. We also {\em assumed} that the 
operator $D$ satisfies  the following relation in the limit 
$a\rightarrow 0$
\begin{eqnarray}
De^{ikx}h(x) &\rightarrow& e^{ikx}(-\kslash+i\delslash
-g\Aslash)h(x)\nonumber\\
&=&i(\delslash+ig\Aslash)(e^{ikx}h(x))
\equiv i\Dslash(e^{ikx}h(x))
\end{eqnarray}
for any {\em fixed} $k_{\mu}$, ($-\frac{\pi}{2a}< k_{\mu}<
\frac{\pi}{2a}$), and a sufficiently smooth function $h(x)$. The 
function $h(x)$ corresponds to the gauge potential in our case, 
which in turn means that the gauge potential $A_{\mu}(x)$
is assumed to vary very little over the distances of the 
elementary lattice spacing. 

Our final expression (3.14) in the limit $M\rightarrow\infty$ 
reproduces the Pontryagin number in the continuum formulation 
\begin{eqnarray}
&&\lim_{M\rightarrow\infty}tr\{ \gamma_{5}f(\Dslash^{2}/M^{2})\}
\nonumber\\
&=& \lim_{M\rightarrow\infty}tr \int \frac{d^{4}k}{(2\pi)^{4}}
e^{-ikx}\gamma_{5}f(\Dslash^{2}/M^{2})e^{ikx}\nonumber\\
&=&\lim_{M\rightarrow\infty}tr \int \frac{d^{4}k}{(2\pi)^{4}}
\gamma_{5}f\{
(ik_{\mu}+ D_{\mu})^{2}/M^{2}+\frac{ig}{4}
[\gamma^{\mu},\gamma^{\nu}]F_{\mu\nu}/M^{2}\}\nonumber\\
&=&\lim_{M\rightarrow\infty}tr M^{4}
\int\frac{d^{4}k}{(2\pi)^{4}}\gamma_{5}f\{
(ik_{\mu}+ D_{\mu}/M)^{2}+\frac{ig}{4}
[\gamma^{\mu},\gamma^{\nu}]F_{\mu\nu}/M^{2}\} 
\end{eqnarray}
where the remaining trace stands for  Dirac and Yang-Mills 
indices. We also
used the relation
\begin{equation}
\Dslash^{2}=D_{\mu}D^{\mu}+\frac{ig}{4}
[\gamma^{\mu},\gamma^{\nu}]F_{\mu\nu}
\end{equation}
and the rescaling of the variable $k_{\mu}\rightarrow M k_{\mu}$. 

By noting  $tr \gamma_{5}
=tr \gamma_{5}[\gamma^{\mu}, \gamma^{\nu}]=0$, the above 
expression ( after expansion in powers of $1/M$) is written as 
(with $ \epsilon^{1234}=1$)
\begin{eqnarray}
\lim_{M\rightarrow\infty}tr \gamma_{5}f(\Dslash^{2}/M^{2})
&=&tr \gamma_{5}\frac{1}{2!}\{\frac{ig}{4}
[\gamma^{\mu},\gamma^{\nu}]F_{\mu\nu}\}^{2}
\int\frac{d^{4}k}{(2\pi)^{4}}f^{\prime\prime}(-k_{\mu}k^{\mu})
\nonumber\\
&&=\frac{g^{2}}{32\pi^{2}}tr \epsilon^{\mu\nu\alpha\beta}
F_{\mu\nu}F_{\alpha\beta}
\end{eqnarray}
where we used 
\begin{eqnarray}
\int \frac{d^{4}k}{(2\pi)^{4}}f^{\prime\prime}(-k_{\mu}k^{\mu})&
=&\frac{1}{16\pi^{2}}\int_{0}^{\infty}f^{\prime\prime}(x)xdx
\nonumber\\
&=& \frac{1}{16\pi^{2}}
\end{eqnarray}
with $x= -k_{\mu}k^{\mu}>0$ in our metric. 

When one combines (3.7) and (3.18), one reproduces  the 
Atiyah-Singer index theorem (in continuum $R^{4}$ space)[10][11]. 
We  note that a local version of the index (anomaly) is valid 
for Abelian theory also.
The global index (3.7) as well as a local version of the index 
(3.8) are both independent of the regulator  $f(x)$  provided [5] 
\begin{equation}
f(0) =1, \ \ \ f(\infty)=0,\ \ \ f^{\prime}(x)x|_{x=0}=f^{\prime}
(x)x|_{x=\infty}=0. 
\end{equation}

We have thus established that the lattice index in (3.7) for any 
algebraic relation in (2.1) is related to the Pontryagin index 
in a smooth continuum limit as
\begin{equation}
n_{+}-n_{-}=\int d^{4}x\frac{g^{2}}{32\pi^{2}}tr
\epsilon^{\mu\nu\alpha\beta}F_{\mu\nu}F_{\alpha\beta}
\end{equation} 
by assuming the quite general properties of the basic operator 
$D$ only: The basic relation (2.1) with hermitian $\gamma_{5}D$ 
and the continuum limit property (3.15) 
{\em without} species doubling in the limit $a\rightarrow 0$.
This shows  that  the instanton-related topological property is 
identical for all the algebraic relations in (2.1), and the 
Jacobian factor (3.6) in fact contains the 
correct chiral anomaly. (We are  implicitly assuming that the 
index (3.7) does not change in the process of taking a continuum 
limit.) Our result is naturally consistent with the calculation 
of chiral anomaly by different methods in [1] and [3].  

\section{Explicit example of the lattice Dirac operator with
$k$=1}

We now discuss an explicit construction of the lattice Dirac 
operator which satisfies the generalized algebraic relation
(2.1) with $k=1$, though a generalization to an arbitrary
$k$ is straightforward as is described in Section 5 
later. For this purpose, we first briefly
review the construction of the Neuberger's overlap Dirac 
operator for the ordinary Ginsparg-Wilson relation.

We start with the conventional Wilson fermion operator $D_{W}$
defined by
 \begin{eqnarray}
D_{W}(x,y)&\equiv&i\gamma^{\mu}C_{\mu}(x,y)+B(x,y)-
\frac{1}{a}m_{0}\delta_{x,y},\nonumber\\
C_{\mu}(x,y)&=&\frac{1}{2a}[\delta_{x+\hat{\mu} a,y}
U_{\mu}
(y)-\delta_{x,y+\hat{\mu} a}U^{\dagger}_{\mu}(x)],
\nonumber\\
B(x,y)&=&\frac{r}{2a}\sum_{\mu}[2\delta_{x,y}-
\delta_{y+\hat{\mu} a,x}U_{\mu}^{\dagger}(x)
-\delta_{y,x+\hat{\mu} a}U_{\mu}(y)],
\nonumber\\
U_{\mu}(y)&=& \exp [iagA_{\mu}(y)],
\end{eqnarray}
where we added a constant mass term to $D_{W}$ for later 
convenience. The parameter $r$ stands for the Wilson parameter.
Our matrix convention is that $\gamma^{\mu}$ are anti-hermitian, 
$(\gamma^{\mu})^{\dagger} = - \gamma^{\mu}$, and thus 
$\Cslash\equiv \gamma^{\mu}C_{\mu}(n,m)$ is hermitian
\begin{equation}
\Cslash^{\dagger} = \Cslash.
\end{equation} 
The operator $D$ introduced by Neuberger[2], which satisfies the 
conventional Ginsparg-Wilson relation (1.1), has an explicit 
expression
\begin{equation}
aD=\frac{1}{2}[1+\gamma_{5}\frac{H_{W}}{\sqrt{H_{W}^{2}}}]
=\frac{1}{2}[1+D_{W}\frac{1}{\sqrt{D_{W}^{\dagger}D_{W}}}]
\end{equation}
where $D_{W}=\gamma_{5} H_{W}$ is the Wilson operator defined 
above, and $H_{W}$ is hermitian $H_{W}^{\dagger}=H_{W}$.

The physical meaning of this construction becomes more transparent
if one considers (naive) near continuum configurations specified 
by a small $a$ limit with the parameters $r/a$ and $m_{0}/a$ kept
finite.
We can then approximate the operator $D_{W}$ by[14]
\begin{equation}
D_{W}\simeq i\Dslash + M_{n}
\end{equation}
for each species doubler, where the mass parameters $M_{n}$ 
stand for $M_{0}= - \frac{m_{0}}{a}$ and one of 
\begin{eqnarray}
&&\frac{2r}{a}-\frac{m_{0}}{a},\ \ (4,-1);\ \ \ 
\frac{4r}{a}-\frac{m_{0}}{a},\ \ (6,1)\nonumber\\
&&\frac{6r}{a}-\frac{m_{0}}{a},\ \ (4,-1);\ \ \ 
\frac{8r}{a}-\frac{m_{0}}{a},\ \ (1,1)
\end{eqnarray}
for $n=1\sim 15$; we denoted ( multiplicity, chiral charge )
 in the bracket for species doublers. Here we used the relation 
valid in the near continuum configurations for the physical 
species, for example, 
\begin{eqnarray}
D_{W}(k) &=& \sum_{\mu}\gamma^{\mu}\frac{\sin ak_{\mu}}{a} + 
\frac{r}{a}\sum_{\mu}(1 - \cos ak_{\mu}) - \frac{m_{0}}{a}
\nonumber\\
&\simeq& \gamma^{\mu}k_{\mu}-\frac{m_{0}}{a}
\end{eqnarray}
in the momentum representation with 
vanishing gauge field. 

In a symbolic notation, one can then write the overlap Dirac 
operator as   
\begin{eqnarray}
aD&\simeq& \sum_{n=0}^{15}\frac{1}{2}[1+(i\Dslash+M_{n})\frac{1}
{\sqrt{\Dslash^{2}+M_{n}^{2}}}]|n\rangle\langle n|,\nonumber\\
a\gamma_{5}D&\simeq& \sum_{n=0}^{15}(-1)^{n}\gamma_{5}
\frac{1}{2}[1+(i\Dslash + M_{n})\frac{1}
{\sqrt{\Dslash^{2}+M_{n}^{2}}}]|n\rangle\langle n|.
\end{eqnarray}
Here we explicitly write the projection  $|n\rangle\langle n|$ 
for each species doubler.
If one chooses the mass parameters so that 
\begin{equation}
M_{0}=-\frac{m_{0}}{a}<0, \ \ \ 
M_{n}>0 \ \ \ for \ \ \ n\neq0
\end{equation}
namely
\begin{equation}
0<m_{0}<2r
\end{equation}
and if one lets all the mass parameters $|M_{n}|$ become large, 
one obtains
\begin{eqnarray}
a\gamma_{5}D&\simeq&\gamma_{5}
\frac{1}{2}[\frac{i\Dslash}{|M_{0}|}+\frac{1}{2}
\frac{\Dslash^{2}}{M_{0}^{2}}]\ \ for\ \ n=0,\nonumber\\
a\gamma_{5}D&\simeq&(-1)^{n}\gamma_{5}
\frac{1}{2}[2+\frac{i\Dslash}{M_{n}}-\frac{1}{2}\frac{\Dslash^{2}}
{M_{n}^{2}}]\ \ for\ \ n\neq0.
\end{eqnarray}
If one chooses $m_{0}$ to satisfy 
\begin{equation}
2a|M_{0}|=2m_{0}=1
\end{equation}
one recovers  the correctly normalized continuum Dirac operator
for the physical species and $\gamma_{5}D\simeq(-1)^{n}\gamma_{5}
\frac{1}{a}$ for unphysical species doublers. 
In particular, the first relation in (4.10) can then be written as
\begin{equation}
H\equiv a\gamma_{5}D\simeq \gamma_{5}ai\Dslash+\gamma_{5}
(\gamma_{5}ai\Dslash)^{2}
\end{equation}
which ensures the conventional Ginsparg-Wilson relation in the 
leading order. These properties become important in the 
following discussion.

\subsection{Generalized algebra with $k=1$}

We now come back to the generalized algebra (2.1) with $k=1$
\begin{equation}
H\gamma_{5}+\gamma_{5}H=2H^{4}
\end{equation}
where $H=a\gamma_{5}D$ and $\Gamma_{5}=\gamma_{5}-H^{3}$.
This algebraic relation implies that
\begin{equation}
\gamma_{5}H^{2}=[\gamma_{5}H+H\gamma_{5}]H-
H[\gamma_{5}H+H\gamma_{5}]+H^{2}\gamma_{5}=H^{2}\gamma_{5}
\end{equation}
Namely, the algebraic relation (4.13) is equivalent to the two 
relations
\begin{eqnarray}
&&H^{3}\gamma_{5}+\gamma_{5}H^{3}=2H^{6},\nonumber\\
&&\gamma_{5}H^{2}-H^{2}\gamma_{5}=0.
\end{eqnarray}
If one defines $H_{(3)}\equiv H^{3}$, the first relation of 
(4.15) becomes 
\begin{equation}
H_{(3)}\gamma_{5}+\gamma_{5}H_{(3)}=2H_{(3)}^{2}
\end{equation}
with $\Gamma_{5}=\gamma_{5}-H_{(3)}$, 
which is identical to the conventional Ginsparg-Wilson relation
(1.1). We utilize this property to construct a solution to 
(4.15). Note that the operator $\Gamma_{5}$ is identical in these
three ways of writing in (4.13), (4.15), and (4.16).

The physical condition for the operator $H$ in (4.13) in the 
near continuum configuration is (Cf.(4.12))
\begin{equation}
H\simeq \gamma_{5}ai\Dslash+\gamma_{5}(\gamma_{5}ai\Dslash)^{4}
\end{equation}
and thus $H_{(3)}$ in (4.16) should satisfy
\begin{eqnarray}
H_{(3)}&\simeq&[\gamma_{5}ai\Dslash
+\gamma_{5}(\gamma_{5}ai\Dslash)^{4}]^{3}\nonumber\\
&\simeq&(\gamma_{5}ai\Dslash)^{3}
+\gamma_{5}(\gamma_{5}ai\Dslash)^{6}
\end{eqnarray}
as can be confirmed by noting $\gamma_{5}\Dslash
+\Dslash\gamma_{5}=0$. Here only the leading terms in chiral 
symmetric and chiral symmetry breaking terms respectively are 
written.

One can thus construct a solution for $H_{(3)}$ by
\begin{equation}
H_{(3)}=\frac{1}{2}\gamma_{5}[1+D_{W}^{(3)}\frac{1}
{\sqrt{(D_{W}^{(3)})^{\dagger}D_{W}^{(3)}}}]
\end{equation}
where we defined $D_{W}^{(3)}$ by\footnote{It is also possible
to use $D_{W}^{(3)}\equiv i(\Cslash)^{3}+(B-\frac{m_{0}}{a})^{3}$,
or any suitable (ultra-local) operator which satisfies 
$\gamma_{5}D_{W}^{(3)}=(\gamma_{5}D_{W}^{(3)})^{\dagger}$ and 
(4.22).}
\begin{equation}
D_{W}^{(3)}\equiv i(\Cslash)^{3}+(B)^{3}-(\frac{m_{0}}{a})^{3}
\end{equation}
The operators $\Cslash$,$B$ and the parameter $m_{0}/a$ are the 
same as in the original Wilson fermion operator (4.1). By 
rewriting (4.19) as 
\begin{equation}
H_{(3)}=\frac{1}{2}\gamma_{5}[1+\gamma_{5}H_{W}^{(3)}
\frac{1}{\sqrt{H_{W}^{(3)}H_{W}^{(3)}}}]
\end{equation}
in terms of the hermitian $H_{W}^{(3)}\equiv\gamma_{5}
D_{W}^{(3)}=(H_{W}^{(3)})^{\dagger}$ and comparing it with (4.3),
one can confirm that our operator $H_{(3)}$ satisfies the 
relation (4.16). The condition (4.18) is satisfied by noting
\begin{equation}
D_{W}^{(3)}\simeq i(\Dslash)^{3}+(M_{n}^{(3)})^{3}
\end{equation}
in the near continuum configuration, where the mass parameters 
are given by
\begin{eqnarray}
(M_{0}^{(3)})^{3}&\equiv&-(\frac{m_{0}}{a})^{3}\nonumber\\
(M_{n}^{(3)})^{3}&\equiv&\{(\frac{2r}{a})^{3}-
(\frac{m_{0}}{a})^{3},
(\frac{4r}{a})^{3}-(\frac{m_{0}}{a})^{3},
(\frac{6r}{a})^{3}-(\frac{m_{0}}{a})^{3},
(\frac{8r}{a})^{3}-(\frac{m_{0}}{a})^{3}\}\nonumber\\
&&\ \ \ for \ \ \ n\neq0.
\end{eqnarray}
Although we have the same condition on the parameters as 
before
\begin{equation}
0<m_{0}<2r
\end{equation}
to avoid the species doublers, the value of $m_{0}$ itself is 
now required to satisfy
\begin{equation}
2(m_{0})^{3}=1
\end{equation}
to ensure the properly normalized physical condition (4.18).

\subsection{Reconstruction of $H$ from $H_{(3)}$}

We now discuss how to reconstruct $H$, which satisfies (4.13),
 from  $H_{(3)}$ defined
above. The basic idea is to take a real cubic root of $H_{(3)}$
as
\begin{equation}
H=(H_{(3)})^{1/3}
\end{equation}
in such a manner that $H$ thus obtained satisfies the second 
constraint in (4.15). For this purpose, we first recall the 
essence of the general representation of the algebra (2.1) 
analyzed in Section 2, which is applicable to (4.16) as well.\\

If one defines the eigenvalue problem
\begin{equation}
H_{(3)}\phi_{n}=(a\lambda_{n})^{3}\phi_{n},\ \ \ 
(\phi_{n},\phi_{n})=1
\end{equation}
one can classify the eigenstates into the 3 classes:\\
(i)\ $n_{\pm}$ (``zero modes''),\\
\begin{equation}
H_{(3)}\phi_{n}=0, \ \ \gamma_{5}\phi_{n} = \pm \phi_{n},
\end{equation}
(ii)\ $N_{\pm}$ (``highest states''), \\
\begin{equation}
H_{(3)}\phi_{n}= \pm\phi_{n}, \ \ \
\gamma_{5}\phi_{n} = \pm \phi_{n},\ \ \ respectively,
\end{equation}
(iii)``paired states'' with $0 < |(a\lambda_{n})^{3}| < 1$,
\begin{equation}
H_{(3)}\phi_{n}= (a\lambda_{n})^{3}\phi_{n}, \ \ \ 
H_{(3)}(\Gamma_{5}\phi_{n})
= - (a\lambda_{n})^{3}(\Gamma_{5}\phi_{n}).
\end{equation}
where
\begin{equation}
\Gamma_{5}=\gamma_{5}-H_{(3)}.
\end{equation}
Note that $\Gamma_{5}(\Gamma_{5}\phi_{n})\propto \phi_{n}$ for 
$0<|(a\lambda_{n})^{3}|<1$.\\

We obtain the index relation
\begin{eqnarray}
Tr\Gamma_{5}&\equiv& \sum_{n}(\phi_{n},\Gamma_{5}\phi_{n})
\nonumber\\
&=&\sum_{\lambda_{n}=0}(\phi_{n},\gamma_{5}\phi_{n})
\nonumber\\
&=& n_{+} - n_{-} =  index
\end{eqnarray}
where $n_{\pm}$ stand for the number of  normalizable zero modes
in the classification (i) above.

We also have a chirality sum rule
\begin{equation}
n_{+}+N_{+}=n_{-}+N_{-}  
\end{equation} 
where $N_{\pm}$ stand for the number of ``highest states''
in the classification (ii) above.

If one denotes the number of states  in the 
classification (iii) above by $2N_{0}$, the total number of 
states (the dimension of the representation) $N$ is given by 
\begin{equation}
N = 2(n_{+}+N_{+}+N_{0})
\end{equation}
which is expected to be common to all the fermion operators 
defined on the same lattice.

Also, all the states $\phi_{n}$ with 
$0<|(a\lambda_{n})^{3}|<1$, 
which appear pairwise with $(a\lambda_{n})^{3}= \pm 
|(a\lambda_{n})^{3}|$, 
can be normalized to satisfy the relations
\begin{eqnarray}
\Gamma_{5}\phi_{n}&=&
[1-(a\lambda_{n})^{6}]^{1/2}\phi_{-n},
\nonumber\\
\gamma_{5}\phi_{n}&=&(a\lambda_{n})^{3}\phi_{n}+
[1-(a\lambda_{n})^{6}]^{1/2}\phi_{-n},
\end{eqnarray}
where $\phi_{-n}$ stands for the eigenstate with an eigenvalue
opposite to that of $\phi_{n}$.\\

Based on these general results in Section 2, we first observe 
that the index $n_{+} - n_{-}$ in (4.32) is identical to the 
index of the expected solution of (4.13), although $H_{(3)}$ 
satisfies (4.18). This observation is 
based on the relation
\begin{equation}
n_{+} - n_{-}\equiv\sum_{n}(\phi_{n},\Gamma_{5}
f((H_{(3)})^{2}/(aM)^{6})\phi_{n})
\end{equation}
which is valid for any regulator with $f(0)=1$. One can perform 
the same analysis as in (3.7) in Section 3: The basic ingredient
is the condition (4.18) for a physical 
momentum region in the smooth continuum limit  and the absence 
of species doublers. 
The calculation analogous to (3.14) then gives
\begin{equation}
n_{+}-n_{-}=\lim_{M\rightarrow\infty}Tr\gamma_{5}
f(\frac{\Dslash^{6}}{M^{6}})
=\lim_{M\rightarrow\infty}Tr\gamma_{5}
g(\frac{\Dslash^{2}}{M^{2}})
\end{equation}
with $g(x)\equiv f(x^{3})$ and $g(0)=1$. The right-hand side 
of this relation shows that the present 
index is identical to the index of the general operator in 
(2.1), which includes an expected solution of (4.13).
Due to the chirality sum rule (4.33), we also obtain the same 
value of $N_{+}-N_{-}$ as for an expected solution of (4.13).\\

The agreement of  the index of $H_{(3)}$ with the index of
the expected solution $H$ of (4.13) suggests that we can define 
$H$ {\em operationally} by
\begin{equation}
H\phi_{n}\equiv a\lambda_{n}\phi_{n}
\end{equation}
by using the {\em same set} of eigenfunctions and (the cubic
roots of) eigenvalues
\begin{equation}
\{\phi_{n}\}, \ \ \ \ \{a\lambda_{n}\}
\end{equation}
as for $H_{(3)}$ in (4.27). Note that the operator 
$\Gamma_{5}=\gamma_{5}-H_{(3)}=\gamma_{5}-H^{3}$, which reverses 
the signature of eigenvalues of ``paired states'' and  
defines the index, is consistently chosen to be identical for 
(4.16) and for 
(4.38)\footnote{This means that an explicit  calculation of 
the chiral Jacobian (and chiral anomaly) for the theory defined 
by (4.13) is performed by $Tr\Gamma_{5}=Tr(\gamma_{5}-
H_{(3)})$ in terms of $H_{(3)}$ in (4.19).}. 

We can then confirm the second 
constraint  in (4.15) and the defining algebraic relation (4.13) 
for {\em any} ``paired state'' $\phi_{n}$,
\begin{eqnarray}
[H^{2}\gamma_{5}-\gamma_{5}H^{2}]\phi_{n}
&=&H^{2}\gamma_{5}\phi_{n}-\gamma_{5}(a\lambda_{n})^{2}\phi_{n}
\nonumber\\
&=&H^{2}\{(a\lambda_{n})^{3}\phi_{n}+
[1-(a\lambda_{n})^{6}]^{1/2}\phi_{-n}\}\nonumber\\
&&-(a\lambda_{n})^{2}\{(a\lambda_{n})^{3}\phi_{n}+
[1-(a\lambda_{n})^{6}]^{1/2}\phi_{-n}\}\nonumber\\
&=&0
\end{eqnarray}
and
\begin{equation}
[\Gamma_{5}H+H\Gamma_{5}]\phi_{n}=
\Gamma_{5}(a\lambda_{n})\phi_{n}-
a\lambda_{n}(\Gamma_{5}\phi_{n})=0
\end{equation}
where we used the relations in (4.35) and the definition (4.38). 
For ``zero modes'' and the ``highest states'', which are the 
eigenstates of $\gamma_{5}$, the condition 
$[H^{2}\gamma_{5}-\gamma_{5}H^{2}]\phi_{n}=0$ obviously holds,
and the relation $[\Gamma_{5}H+H\Gamma_{5}]\phi_{n}=0$ is also
confirmed.

The general representation of the algebra (4.13) is obtained 
from the {\em standard representation}, which is defined by $H$ 
in (4.38), $\gamma_{5}$ in (4.35), and the state vectors 
$\{\phi_{n}\}$ in (4.39), by applying a suitable unitary 
transformation.

\section{Discussion}

When one considers the algebraic relation with a constant $R$
\begin{equation}
\gamma_{5}(\gamma_{5}D)+(\gamma_{5}D)\gamma_{5}=2Ra^{2k+1}
(\gamma_{5}D)^{2k+2}
\end{equation}
instead of (2.1), one can eliminate the paprameter $R$ by a 
scale transformation
\begin{equation}
D\rightarrow D^{\prime}=R^{1/(2k+1)}D.
\end{equation}
The path integral 
\begin{equation}
\int{\cal D}\bar{\psi}{\cal D}\psi
\exp[\int\bar{\psi}D^{\prime}\psi]
\end{equation}
is equivalent to 
\begin{equation}
\int{\cal D}\bar{\psi}{\cal D}\psi\exp[\int\bar{\psi}D\psi]
\end{equation}
after absorbing the parameter $R^{1/(2k+1)}$ into $\bar{\psi}$,
at least in a well regularized lattice path integral.
Consequently, the parameter $R$ and also the factor $a^{2k+1}$ 
do not have an intrinsic physical significance\footnote{
However, when one includes a Yukawa interaction, for example, 
this scaling argument need to be refined.}.

In contrast, the power of $(\gamma_{5}D)^{2k+2}$ in the right-hand
side of (5.1) has an intrinsic physical meaning. One may recall
the near continuum expressions (4.12) and (4.17)
\begin{eqnarray}
H&\simeq&\gamma_{5}ai\Dslash+\gamma_{5}(\gamma_{5}ai\Dslash)^{2}
\ \ for\ \ k=0,\nonumber\\
H&\simeq&\gamma_{5}ai\Dslash+\gamma_{5}(\gamma_{5}ai\Dslash)^{4}
\ \ for\ \ k=1
\end{eqnarray}
respectively. The first terms in these expressions stand for the 
leading terms in chiral symmetric terms, and the second terms in
these expressions stand for the leading terms in chiral symmetry
breaking terms. This shows that one can improve the chiral 
symmetry\footnote{To avoid the mis-understanding, we note that 
the improvement of chiral symmetry here is meant in the sense of 
Wilsonian renormalization group. The chiral symmetry breaking 
term becomes more irrelevant for larger $k$, and this should be 
interesting from a view point of regularization of field theory 
in general. Also, the approach to the 
continuum Dirac operator is controlled by two parameters, for
example, by letting $k\rightarrow large$ and 
$a\rightarrow small$ simultaneously.} 
by choosing a large parameter $k$. 

The Dirac operator for such a general value of $k$ is constructed 
by rewriting (2.1) as a set of relations (see (4.14))
\begin{eqnarray}
&&H^{2k+1}\gamma_{5}+\gamma_{5}H^{2k+1}=2H^{2(2k+1)},\nonumber\\
&&H^{2}\gamma_{5}-\gamma_{5}H^{2}=0,
\end{eqnarray}
with $H=a\gamma_{5}D$. The first of these relations (5.6) becomes
identical to the ordinary Ginsparg-Wilson relation (1.1) if one 
defines
$H_{(2k+1)}\equiv H^{2k+1}$. 
One can construct a solution to (5.6) by following the 
prescription in Section 4 
\begin{equation}
H_{(2k+1)}=\frac{1}{2}\gamma_{5}[1+D_{W}^{(2k+1)}\frac{1}
{\sqrt{(D_{W}^{(2k+1)})^{\dagger}D_{W}^{(2k+1)}}}]
\end{equation} 
where 
\begin{equation}
D_{W}^{(2k+1)}\equiv i(\Cslash)^{2k+1}+B^{2k+1}
-(\frac{m_{0}}{a})^{2k+1}
\end{equation}
The operator $H$ is then finally defined by (in the 
representation where $H_{(2k+1)}$ is diagonal)
\begin{equation}
H=(H_{(2k+1)})^{1/2k+1}
\end{equation}
in such a manner that the second relation of (5.6) is satisfied.
This condition is in deed satisfied as a generalization of (4.40)
 in the representation where $H_{(2k+1)}$ is diagonal. We use the 
relation (2.23) in this proof. Also the conditions $0<m_{0}<2r$
and 
\begin{equation}
2m_{0}^{2k+1}=1
\end{equation}
ensure a proper normalization of the Dirac
operator $H$.

However, one need to use a large enough lattice to accomodate 
the operator $H$ with a large $k$, since the operator (5.8) 
correlates lattice points far apart from each other for a large 
$k$. An explicit analysis of the locality property of our 
operator $H$ as in Ref.[15] is left as an important problem. 
In the context of lattice simulation, it would  be 
interesting to see how the chiral properties are modified if one 
uses the operator with $k=1$, which has been analyzed in detail 
in this paper, instead of the conventional overlap Dirac 
operator with $k=0$. To detect the possible effects of $k\neq0$
in a reliable way, it is expected that one would have to 
consider a sufficiently large lattice and those observables 
which are sensitive to low energy excitations.

As for the chiral fermions on the lattice, our general
algebra (2.1) satisfies the decomposition
\begin{equation}
D=\frac{(1+\gamma_{5})}{2}D\frac{(1-\hat{\gamma}_{5})}{2}
+\frac{(1-\gamma_{5})}{2}D\frac{(1+\hat{\gamma}_{5})}{2}
\end{equation}
with
\begin{equation}
\hat{\gamma}_{5}\equiv\gamma_{5}-2a^{2k+1}(\gamma_{5}D)^{2k+1},
\ \ \ \ \  (\hat{\gamma}_{5})^{2}=1
\end{equation}
by noting 
$\gamma_{5}(\gamma_{5}D)^{2}=(\gamma_{5}D)^{2}\gamma_{5}$. 
This decomposition has the same form as for the overlap 
operator $D$ satisfying the ordinary Ginsparg-Wilson relation.
It is thus expected that one can apply the same considerations
as in Refs.[16] and [17] to our general Dirac operator also.
In particular, the fermion number non-conservation of the 
chiral theory defined by 
\begin{eqnarray}
&&\int{\cal D}\bar{\psi}{\cal D}\psi
\exp\{\int\bar{\psi}D_{L}\psi\}\nonumber\\
&&\equiv\int{\cal D}\bar{\psi}{\cal D}\psi\exp\{\int\bar{\psi}
\frac{(1+\gamma_{5})}{2}D\frac{(1-\hat{\gamma}_{5})}{2}\psi\}
\end{eqnarray}
follows from the fermion number transformation
\begin{equation}
\psi\rightarrow e^{i\alpha}\psi,\ \ \ \bar{\psi}\rightarrow
\bar{\psi}e^{-i\alpha}.
\end{equation}
If one remembers that the functional spaces of the variables 
$\psi$ and $\bar{\psi}$ are specified by the 
projection operators $(1-\hat{\gamma}_{5})/2$ and 
$(1+\gamma_{5})/2$, respectively, the Jacobian factor 
for the transformation (5.14) is given by[16]
\begin{eqnarray}
J&=&\exp \{i\alpha Tr[\frac{(1+\gamma_{5})}{2}-
\frac{(1-\hat{\gamma}_{5})}{2}]\}\nonumber\\
&=&\exp\{i\alpha Tr{}[\gamma_{5}-(\gamma_{5}aD)^{2k+1}{]}  \} 
=\exp\{i\alpha{[}n_{+}-n_{-}{]}\}
\end{eqnarray}
where the index is defined in (2.19).

In conclusion, we have shown that the general idea of 
Ginsparg and Wilson can be precisely realized as a closed 
algebraic relation (2.1) and it admits of the explicit 
construction of an infinite tower of new lattice Dirac operators 
as a generalization of the overlap Dirac operator. This should be 
interesting in the context of the regularization of field theory 
in general. 
\\
\\
\\
{\bf Acknowledgement}\\
\\
The present work was initiated when I was visiting at Center
for Subatomic Structure of Matter(CSSM), University of Adelaide.
I am grateful to David Adams and T-W. Chiu for stimulating 
discussions, and to Anthony Williams and David Adams for their 
hospitality at CSSM.

\end{document}